\documentclass[10pt,letterpaper]{article}
\usepackage{opex3}
\usepackage{hangcaption}
\usepackage{array}
\usepackage{indentfirst} 
\usepackage{cite}

\begin{document}

%% NOTE: TITLE PAGE & TOC NOT USED FOR MANUSCRIPT SUBMISSIONS %%
%\title{Real-time digital holographic microscopy using the graphic processing unit}

%\vskip4pc

%\tableofcontents
%\clearpage
%% NO TITLE PAGE FOR OPEX SUBMISSIONS %%

%% START HERE
%%%%%%%%%%%%%%%%%% title page information %%%%%%%%%%%%%%%%%%
\title{Optical encryption for large-sized images using random phase-free method}

\author{
Tomoyoshi Shimobaba,$^{1*}$
Takashi Kakue,$^1$
Yutaka Endo,$^{1}$\\
Ryuji Hirayama,$^{1}$
Daisuke Hiyama,$^{1}$
Satoki Hasegawa,$^{1}$\\
Yuki Nagahama,$^{1}$
Marie Sano,$^{1}$
Takashige Sugie,$^{1}$
and 
Tomoyoshi Ito$^1$
}

\address{$^1$ Graduate School of Engineering, Chiba University, 1-33 Yayoi-cho, Inage-ku, Chiba 263-8522, Japan}

\begin{abstract} 
We propose an optical encryption framework that can encrypt and decrypt large-sized images beyond the size of the encrypted image using our two methods: random phase-free method and scaled diffraction.
In order to record the entire image information on the encrypted image, the large-sized images require the random phase to widely diffuse the object light over the encrypted image; however, the random phase gives rise to the speckle noise on the decrypted images, and it may be difficult to recognize the decrypted images.
In order to reduce the speckle noise, we apply our random phase-free method to the framework.
In addition, we employ scaled diffraction that calculates light propagation between planes with different sizes by changing the sampling rates.
\end{abstract}

 % REPLACE WITH CORRECT OCIS CODES FOR YOUR ARTICLE

\

%%%%%%%%%%%%%%%%%%%%%%% References %%%%%%%%%%%%%%%%%%%%%%%%%

\section{Introduction}
\noindent Optical encryption methods \cite{review1,review2} are promising technology because they have unique properties of parallel processing ability and multiple keys, which are random phase, wavelength, polarization and propagation distance. 
Double random phase encoding (DRPE) \cite{drp} is pioneering work, in which an original image is encrypted by using two random phase planes and then recorded as a hologram.
Subsequently to this pioneering work, in order to reinforce the decoding resistance, a large number of optical encryption schemes have been proposed: for example, the extension to Fresnel domain \cite{drp_fre}, binary key code \cite{nomura}, fractional Fourier transform-based \cite{fractional},  gyrator transform-based \cite{gyrator}, M-sequence \cite{tsang}, and ghost imaging-based schemes \cite{ghost}.
Optical encryption is also suitable for optical memory because of writing and reading data from the optical memory as encrypting and decrypting the data \cite{matoba}.

In the decryption process in optical encryption, the speckle noise in decrypted images is often a problem and it hinders the recognition of the decrypted information.
Recently, optical encryption methods using Quick Response (QR) code tolerant to the speckle noise \cite{qr1,qr2} have been proposed. 
In these methods, the original information is first converted to QR code and then the QR code is encrypted. 
Even if the decrypted QR code is contaminated by noise, the information can be retrieved due to the property of the QR code.

If we want to encrypt large-sized images beyond the size of the encrypted image, we need to apply random phase, which is one of the causes of the speckle noise, to the images because it is necessary to widely diffuse the light of the images. 
In order to avoid degradation of the decrypted images, we can use the QR code schemes; unfortunately, the QR code does not contain a large amount of information.

In this paper, we propose an optical encryption framework that can encrypt and decrypt large-sized images beyond the size of the encrypted image using  two methods: random phase-free method \cite{random_free} and scaled diffraction \cite{shift,arss}.
Recently, we proposed a random phase-free method and it can record and reconstruct large-sized images without using the random phase.
This method can dramatically reduce the speckle noise.
Scaled diffraction calculates the light propagation between planes with  different sizes by changing the sampling rates on the original and encrypted images.
Section 2 describes our optical encryption framework for large-sized images, Section 3 shows the simulation results of the proposed method and Section 4 concludes this work.

\section{Optical encryption framework for large-sized images}

\begin{figure}[htb]
\centerline{\includegraphics[width=10cm]{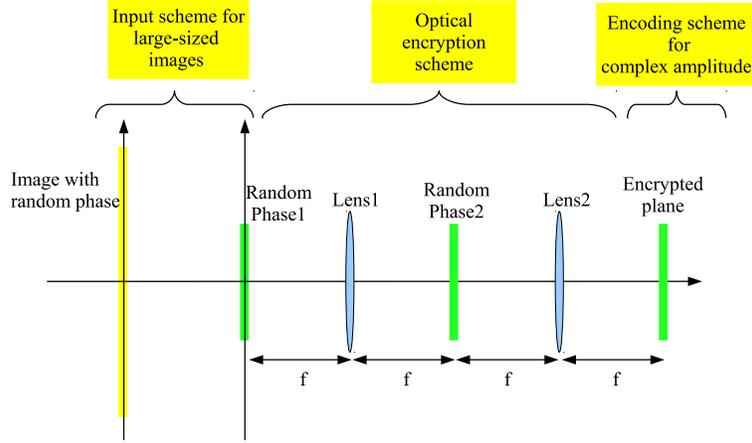}}
\caption{Optical encryption framework for large-sized images with random phase.}
\label{fig:sys_rand}
\end{figure}

Figure \ref{fig:sys_rand} shows the optical encryption framework for large-sized images.
The framework consists of ``Input scheme'' for large-sized images, ``Optical encryption scheme'' and ``Encoding scheme'' for complex amplitude.
The input scheme plays the role of inputting the light from large-sized images to the encryption scheme with as little degradation as possible.
The encryption scheme encrypts the light from the input scheme, e.g. we can use DRPE, Fresnel-base DRPE, fractional Fourier transform-based and gyrator transform-based optical encryption.
The last part, the encoding scheme, encodes the complex amplitude from the output of the encryption scheme to a hologram, e.g. we can use digital holography \cite{dh1,dh2}. 
Figure \ref{fig:sys_rand} shows a conventional method for encrypting large-sized images. 
In this conventional method, we apply the random phase to the input image in the input scheme, and we use DRPE as the optical encryption scheme.

Most images in general contain many low frequency components, and thus cannot spread light widely because the spread angle is proportional to $\sin^{-1}(\lambda \nu)$ where $\lambda$ and $\nu$ are the wavelength and the spatial frequency of the image, respectively \cite{random_free}.
In the case without the random phase on the large-sized images, the low frequency parts on the images cannot reach the first random phase in DRPE; therefore, the parts cannot be encrypted.
The random phase on the images is required for widely spreading the light.  

\begin{figure}[htb]
\centerline{\includegraphics[width=10cm]{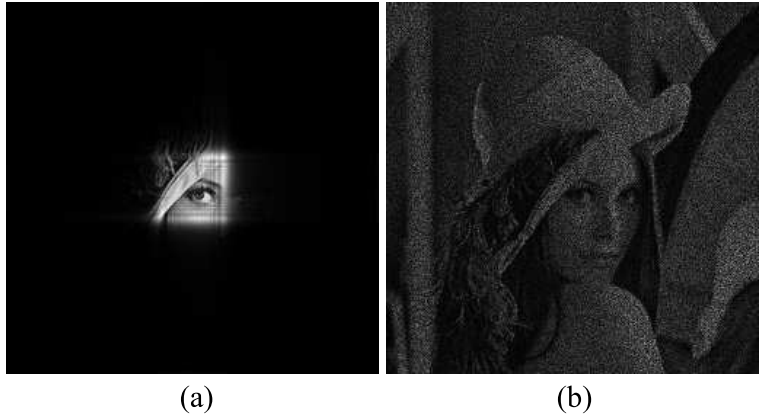}}
\caption{Simulation results of decrypted images. (a) without the random phase on the input image (b) with the random phase on the input image.}
\label{fig:lena_rand}
\end{figure}

Figure \ref{fig:lena_rand} shows the simulation results of decrypted images with and without the random phase on the input image.
The parameters of the simulation are the wavelength of 532 nm, the distance between the images and the first random phase in DRPE of 1.5 m, the resolution of the input image and the encrypted image of $N \times N = 2,048 \times 2,048$ pixels, and the pixel pitches on the input image and the others (Random phases 1, 2 and the encrypted plane) of $p_i=20 \mu$m and $p_o = 4 \mu$m, respectively.
Figure \ref{fig:lena_rand} (a) shows the decrypted image without the random phase on the input image.
As we can see, only part of the image is decrypted, whereas the decrypted image in Fig. \ref{fig:lena_rand} (b) can decrypt the entire image; unfortunately, the decrypted image is contaminated by speckle noise.

\begin{figure}[htb]
\centerline{\includegraphics[width=10cm]{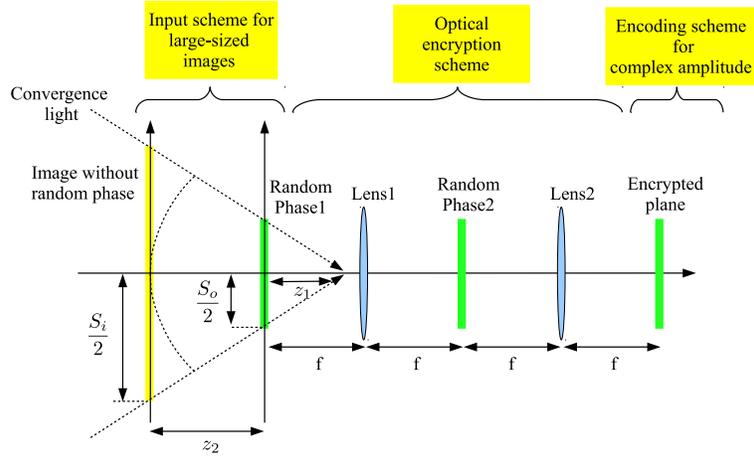}}
\caption{Optical encryption framework for large-sized images with the proposed method.}
\label{fig:sys_propose}
\end{figure}

Figure \ref{fig:sys_propose} shows optical encryption framework for large-sized images using the random phase-free method \cite{random_free}, which can encrypt large-sized images without applying random phase to the images, thus we expect that the reconstructed image is not contaminated by speckle noise.
The only difference between Figs. \ref{fig:sys_rand} and \ref{fig:sys_propose} is the use of spherical convergence light, instead of random phase on the input image.
The areas of the input image and the others are $S_{i} \times S_{i}$ ($S_i =N \times p_i$) and $S_{o} \times S_{o}$ ($S_o =N \times p_o$), respectively.
In the proposed method, we multiply the input image $u_i(x_i, y_i)$ by the convergence light given by $w(x_i, y_i)$.
 $w(x_i, y_i)$ is expressed as a convergence spherical light $w(x_i, y_i)=\exp(-i \pi(x_o^2+y_o^2)/\lambda f_i)$ where $f_i=z_1+z_2$ is the focal length.
The distance between the focus point of the convergence light and the first random phase in DPRE is denoted by $z_1$, and is set to the distance at which Random phase 1 just fits to the cone of the convergence light.
The distance between the input image and the first random phase (this is the decryption distance) is denoted by $z_2$.
Subsequently, we calculate the complex amplitude on Random phase 1 using $u_h(x_h, y_h) = {\rm Prop_{z_2}}\{u_o(x_o, y_o) w(x_o, y_o)\}$ where $ {\rm Prop_{z_2}\{\cdot \}}$ denotes diffraction calculation at the propagation distance $z_2$.
Using a simple geometric relation as shown in Fig.\ref{fig:sys_propose}, we can derive $S_{i}/2 : S_{o}/2 = z_1: f_i$, hence $f_i=z_2/(1-S_{o}/ S_{i})$.
See more details in Ref. \cite{random_free}.

When we use virtual optics-based encryption \cite{virtual1,virtual2,virtual3}, scaled diffraction \cite{shift,arss} that calculates light propagation between planes with different sizes by changing the sampling rates can be used for the encryption and decryption.
In this paper, we used ARSS Fresnel diffraction \cite{arss}, which improved the aliasing problem of Shifted-Fresnel diffraction \cite{shift}. 

\begin{figure}[htb]
\centerline{\includegraphics[width=8cm]{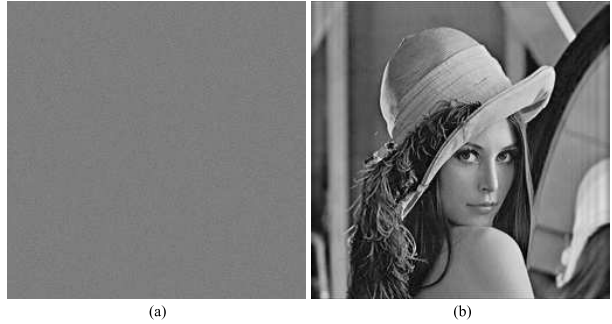}}
\caption{Simulation result using the proposed method. (a) encrypted image (b) decrypted image.}
\label{fig:lena_propose}
\end{figure}

%To avoid the aliasing error, the incident angle $\theta$ of the convergence light must satisfy $\theta=\sin^{-1}(\lambda/2 p_h)$ where $\lambda$ is the wavelength and $p_h$ is the sampling rate on the CGH, respectively.

\section{Results}
Figure \ref{fig:lena_propose} shows the simulation result of the proposed method.
The simulation conditions are the same as those shown in Fig. \ref{fig:lena_rand}.
Figure \ref{fig:lena_propose} (a) shows the encrypted image and (b) shows the decrypted image.
The decrypted image is well improved, compared to Fig. \ref{fig:lena_rand}.
The sizes of the decrypted images (Figs.\ref{fig:lena_rand} and  \ref{fig:lena_propose}) and the encrypted image are approximately 41 mm ($N \times p_i$) and 8.2 mm ($N \times p_o$), respectively.
The peak signal-to-noise ratios (PSNRs) between the original image and the decrypted images of Figs.\ref{fig:lena_rand}(b) and \ref{fig:lena_propose} are 9.1 dB and 29 dB, respectively.

\begin{figure}[htb]
\centerline{\includegraphics[width=8cm]{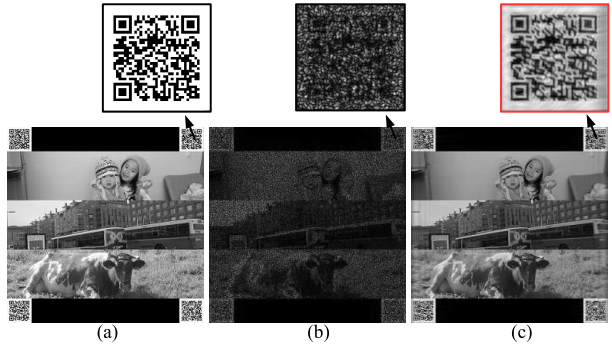}}
\caption{Decrypted results of four QR codes and three images.
(a) Original image including four QR codes and three images. (b) decrypted image from the encrypted image using the conventional method (c) decrypted image from the encrypted image using the proposed method.}
\label{fig:qr1}
\end{figure}

We show other simulation results in Fig. \ref{fig:qr1}.
Figure \ref{fig:qr1}(a) shows the original images including four QR codes and three images.
The upper left QR code corresponds to the Uniform Resource Locator (URL) of ``http://brains.chiba-u.jp/~shimo/''.
The upper right, bottom left and bottom right QR codes correspond to URLs of ``http://www.opticsinfobase.org/'',  ``http://www.sciencemag.org/'' and ``http://www.nature.com/nphoton/index.html/'', respectively. 
Figure \ref{fig:qr1}(b) shows the decrypted image using the conventional method as shown in Fig.\ref{fig:sys_rand}.
Although the images can be barely recognized, none of the decrypted QR codes can decode the URLs by the author's cellular phone.
Whereas, Fig. \ref{fig:qr1}(c) shows the decrypted image using the proposed method as shown in Fig.\ref{fig:sys_propose} and all the decrypted QR codes can decode the URLs correctly by the author's cellular phone because of the low speckle noise.

\begin{figure}[htb]
\centerline{\includegraphics[width=8cm]{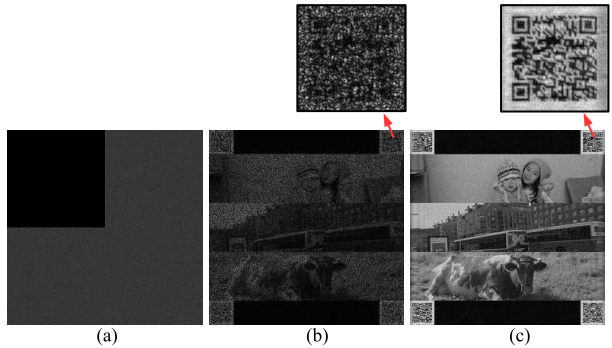}}
\caption{Decrypted results from the encrypted image of which one fourth has vanished.
(a) Broken encrypted image (one fourth of the encrypted image has vanished)  (b) Decrypted image from the encrypted image using the conventional method (c) Decrypted image from the encrypted image using the proposed method.}
\label{fig:qr2}
\end{figure}

Figure \ref{fig:qr2} shows the decrypted images from the encrypted image of which one fourth has vanished.
Figure \ref{fig:qr2}(a) is the encrypted image.
Figure \ref{fig:qr2}(b) shows the decrypted image from the broken encrypted image using the conventional method.
The QR codes and three images are further contaminated by noise, and the QR codes cannot be decoded.
Whereas, although the decrypted image in Fig. \ref{fig:qr2}(c) is slightly contaminated by noise, the QR codes can still be decode the URLs correctly.
The PSNRs between the original image and the decrypted images of Figs.\ref{fig:qr2}(b) and \ref{fig:qr2}(c) are 8.3 dB and 16 dB, respectively.
All of the calculations in this paper were performed by our numerical library for wave optics \cite{cwo}.

\section{Conclusion}
\noindent We proposed an optical encryption framework, that can encrypt and decrypt large-sized images beyond the size of the encrypted image, utilizing our random phase-free method and scaled diffraction.
A large-sized image beyond the encrypted image size requires random phase; unfortunately, it is a factor in the generation of considerable speckle noise on the decrypted images, and it may be hard to decrypt the information even if using QR codes.
We succeeded in improving this problem thanks to our random phase-free method, and reconstructing the large-sized decrypted image by numerical means of scaled diffraction.

\section*{Acknowledgments}
\noindent This work is partially supported by JSPS KAKENHI Grant Numbers 25330125 and 25240015, and the Kayamori Foundation of Information Science Advancement and Yazaki Memorial Foundation for Science and Technology.
\end{document}